\documentclass[aps,prd,a4paper,onecolumn,notitlepage,10pt]{revtex4-1}
\usepackage{amsmath}
\usepackage{bm}
\begin{document}
\title{Gravity with a Dynamical Spinning Aether}
\author{Christopher Kohler}
\altaffiliation{Materials Testing Institute University of
  Stuttgart,\\Pfaffenwaldring 32, 70569 Stuttgart, Germany}
\email[\\E-mail address: ]{christopher.kohler@mpa.uni-stuttgart.de}
\begin{abstract}
Einstein-aether theory is extended by allowing for spinning degrees of
freedom of the aether. In addition to the acceleration, shear,
expansion, and vorticity of the aether velocity field, a spin rotation
describing the dynamics of a classical intrinsic angular momentum of
the aether is introduced as a kinematic quantity. The action of
Einstein-aether theory is augmented by a term quadratic in the spin
rotation and by coupling terms with the vorticity and the
acceleration. Besides breaking the Lorentz boost invariance, the
theory breaks the invariance under spatial rotations in the direction
of the aether velocity. In the weak field limit, there is a linear
relationship between the spin rotation, the vorticity, and the
acceleration. Linearized wave solutions correspond to the ones of
Einstein-aether theory where the speeds of the spin 0 and spin 1 mode
are modified. The extension of Einstein-aether theory has a natural
formulation in the framework of a teleparallel geometry where the
kinematic quantities become torsion fields.
\end{abstract}
\maketitle
\section{Introduction}

Einstein-aether theory is a theory of gravity where a dynamical unit
timelike vector field --- the aether --- is coupled to general
relativity~\cite{jacobson01,eling06,jacobson07}. The vector field can
be regarded as a kind of preferred frame violating local Lorentz
invariance of the theory in that the symmetry under local Lorentz
boosts is broken while the symmetry under local spatial rotations is
retained.  In its original formulation, the interpretation of the
vector field as a four-velocity in Einstein-aether theory is achieved
by a unit norm constraint imposed by a Lagrange multiplier in the
action functional thus implying a spontaneous Lorentz boost symmetry
breaking. If, however, the vector field in the action is implicitly
assumed to be timelike and of unit norm, the approach represents a
semi-tetrad formulation and the symmetry under Lorentz boosts is
explicitly broken.

The timelike vector field can be seen as a fluid existing everywhere
in space-time. In this work, we will consider an extension of
Einstein-aether theory that is motivated by the physics of
spin-fluids~\cite{halbwachs60,ray82}. The description of spin in this
paper is purely classical.  It is considered an intrinsic degree of
freedom that is not quantized. The mathematical formulation of the
spinning aether makes it necessary to introduce a spatial triad with
respect to which the spin is fixed.  In the same way as the aether in
Einstein-aether theory is solely given by its four-velocity, the
spinning aether will additionally be described by its triad
orthonormal to the velocity. Hence, the spinning aether is described
by a tetrad.

The invariance of the theory under local spatial rotations will be
retained as far as possible. For this reason, we avoid the appearance
of derivatives of the triad in the spatial directions in the action
functional. Only derivatives of the triad in the direction of the
aether velocity will be allowed which means that only the invariance
under spatial rotations in this direction will be broken. In this way,
isotropy of space is preserved since there is no preferred spatial
direction.

In a formulation within Riemannian geometry, the aether is not a pure
geometrical object. The geometrical objects of Riemannian geometry ---
the metric, the Christoffel symbols, and the curvature tensor --- are
Lorentz invariant; the Lorentz invariance breaking aether is an
additional degree of freedom. Nevertheless, a pure geometrical
formulation of gravity with an aether is possible using non-Riemannian
geometry. A preferred frame defines a distant parallelism in that
vectors with constant components with respect to the frame are
considered parallel.  This motivates the formulation of theories of
gravity with preferred frames using teleparallel gravity. If the
spatial orientation of the frames are not fixed, the natural geometry
corresponds to a partial parallelization of space-time.

This paper is organized as follows.  In section II, the extended
version of Einstein-aether theory is introduced. The kinematic
quantities of the spinning aether are defined and the action
functional is chosen. The field equations are derived in section III
and it is shown in which way they can be simplified using constraint
equations. Section IV examines the weak field limit of the field
equations. The aether excitations are explored using analogies with
Maxwell's equations. In section V, the extended Einstein-aether theory
is formulated as a (semi-)teleparallel theory of gravity. The
geometry is defined and a procedure to find action functionals is
described.

We use the following conventions: Greek indices $\mu, \nu,\rho, ...$
with the range $0,1,2,3$ denote space-time indices. Latin indices $a,
b, c, ...$ with the range $0,1,2,3$ are internal indices. Spatial
indices in the range $1,2,3$ are denoted by latin indices from the
middle of the alphabet, $i,j,k,...$.  The metric signature is
$(+,-,-,-)$. Symmetrization is denoted by round brackets,
antisymmetrization by square brackets. The totally antisymmetric
pseudotensor is $\varepsilon_{\mu\nu\rho\sigma}$. Units are chosen in
which $c=1$.

%%%%%%%%%%%%%%%%%%%%%%%%%%%%%%%%%%%%%%%%%%%%%%%%%%%%%

\section{Action Functional}

We assume that the preferred frame is given by a tetrad $e_a{}^\mu$
where the four velocity of the aether is $u^\mu = e_0{}^\mu$. The
tetrad is assumed to be orthonormal which means that the space-time
metric is given by $g_{\mu\nu} = e^a{}_\mu e^b{}_\nu \eta_{ab}$ where
$e^a{}_\mu$ is the cobasis defined by $ e_a{}^\mu e^a{}_\nu =
\delta^\mu_\nu$ and $ \eta_{ab} = \textrm{diag}(1,-1,-1,-1)$ is the
Minkowksi metric.

The spatial projection tensor determined by the tetrad is given by
\begin{equation}
  h^\mu_\nu = \delta^\mu_\nu - u^\mu u_\nu = e_i{}^\mu e^{i}{}_\nu.
\end{equation}

The covariant derivative corresponding to the metric $g_{\mu\nu}$ will
be denoted by $D_\mu$. Using the covariant derivative, we can compute
the space-time components of the Ricci rotation
coefficients~\cite{moller61},
\begin{equation}
  C_{\mu\nu\rho} = e^a{}_\mu D_\rho e_{a\nu},
\end{equation}
which measure the deviation of the tetrad from being inertial.

Given the velocity field, kinematic quantities can be defined which
are projections of $C_{\mu\nu\rho}$. The acceleration $a^\mu$ is given
by
\begin{equation}
a^\mu =  D_u u^\mu = u^\rho D_\rho u^\mu = - C^\mu{}_{\nu\rho} u^\nu u^\rho.
\end{equation}
The expansion $\theta$ is defined as
\begin{equation}
  \theta = D_\rho u^\rho = C_{\rho\mu\nu} u^\rho h^{\mu\nu}.
\end{equation}
The shear tensor $\sigma_{\mu\nu}$ being the trace-free part of the
deformation tensor is given by
\begin{align}
  \nonumber
  \sigma_{\mu\nu} & = h^\rho_\mu h^\sigma_\nu D_{(\rho} u_{\sigma)} -
  \frac{1}{3}h_{\mu\nu} \theta = D_{(\mu} u_{\nu)} - u_{(\mu} a_{\nu)}
  - \frac{1}{3}h_{\mu\nu} \theta\\ & = C_{\rho\sigma\tau} u^\rho
  \left( h^\sigma_{(\mu} h^\tau_{\nu)}- \frac{1}{3}h_{\mu\nu}
  h^{\sigma\tau} \right) .
\end{align}
The vorticity tensor $\omega_{\mu\nu}$ is defined as
\begin{align}
  \nonumber \omega_{\mu\nu} & = h^\rho_\mu h^\sigma_\nu D_{[\rho}
    u_{\sigma]} = \partial_{[\mu} u_{\nu]} - u_{[\mu} a_{\nu]}\\ & = -
  C_{\rho\sigma\tau} u^\rho h^\sigma_{[\mu} h^\tau_{\nu]}.
\end{align}
The kinematic quantities associated with the four velocity $u^\mu$ can
be combined in the tensor $D_\mu u_\nu$ which can be irreducibly
decomposed according to
\begin{equation}
  \label{eq:irrdec}
D_\mu u_\nu = u_\mu a_\nu +
\sigma_{\mu\nu} + \omega_{\mu\nu} + \frac{1}{3} h_{\mu\nu}\theta
= C_{\rho\nu\mu} u^\rho.
\end{equation}

As an additional kinematic quantity, associated with the spatial
triad $e_i{}^\mu$, we define the spin rotation $\kappa_{\mu\nu}$ by
\begin{equation}
  \label{eq:spinrot}
  \kappa_{\mu\nu} = e^i{}_\nu\! \stackrel{F}{D}{\!\!}_u e_i{}_\mu
\end{equation}
where $\stackrel{F}{D}{\!\!}_u$ denotes the Fermi derivative in the
direction of $u^\mu$ given by
\begin{equation}
  \label{eq:fermi}
   \stackrel{F}{D}{\!\!}_u X^\mu = D_u X^\mu + X^\rho a_\rho u^\mu -
   X^\rho u_\rho a^\mu
\end{equation}
for a vector field $X^\mu$. The spin rotation can then be written as
\begin{equation}\label{eq:spinrot2}
  \kappa_{\mu\nu} = e^i{}_\nu D_u e_i{}_\mu + u_\mu a_\nu =
  - C_{\sigma\tau\rho} h^\sigma_{[\mu} h^\tau_{\nu]} u^\rho .
\end{equation}
  If the aether spin vector is assumed to be
fixed with respect to the triad, the spin rotation measures the
deviation (a spatial rotation) of the spin from a Fermi-Walker
transported spin in the direction of the aether velocity.

In order to understand the role of the spin rotation $\kappa_{\mu\nu}$
as a kinematic quantity, we note that the tensor $C_{\mu\nu\rho}$ can
be decomposed according to
\begin{equation}
  \label{eq:decomp}
  C_{\mu\nu\rho} = Q_{\mu\nu\rho} + S_{\mu\nu\rho}
\end{equation}
into its spatial projection
\begin{equation}
  Q_{\mu\nu\rho} = h^\lambda_\mu h^\sigma_\nu h^\tau_\rho C_{\lambda\sigma\tau}
  = h^\sigma_\nu h^\tau_\rho e^i{}_\mu D_\tau e_{i\sigma}
\end{equation}
and its time-space components $S_{\mu\nu\rho}$. Equation
\eqref{eq:decomp} can be solved for $S_{\mu\nu\rho}$ yielding
\begin{align}
  \nonumber
  S_{\mu\nu\rho} & = u_\mu D_\rho u_\nu - u_\nu D_\rho u_\mu -
  \kappa_{\mu\nu} u_\rho\\
  \label{eq:stensor}
  & = 2 u_{[\mu} a_{\nu]} u_\rho + \frac{2}{3}
  u_{[\mu} h_{\nu]\rho} \theta + 2 u_{[\mu} \sigma_{\nu]\rho} - 2
  u_{[\mu} \omega_{\nu]\rho} - \kappa_{\mu\nu} u_\rho.
\end{align}
The tensor $S_{\mu\nu\rho}$ can be seen as the generalization of the
expression $D_\mu u_\nu$ in Equation \eqref{eq:irrdec}. Equation
\eqref{eq:stensor} corresponds to its irreducible decomposition where
the spin rotation $\kappa_{\mu\nu}$ naturally appears as an
irreducible part.

The vorticity tensor and the spin rotation can be represented by
spatial vectors $\omega_\mu$ and $\kappa_\mu$ defined by
\begin{equation}
  \omega_\mu = \frac{1}{2} \varepsilon_{\mu\nu\rho} \omega^{\nu\rho}
\end{equation}
and a similar equation for $\kappa_\mu$. Here, the totally
antisymmetric spatial pseudotensor $\varepsilon_{\mu\nu\rho}$ is given
by $\varepsilon_{\mu\nu\rho} = \varepsilon_{\mu\nu\rho\sigma}
u^\sigma$. We will also introduce an antisymmetric acceleration
tensor $a_{\mu\nu}$ defined by
\begin{equation}
  a_{\mu\nu} = - \varepsilon_{\mu\nu\rho} a^\rho.
\end{equation}

In Einstein-Aether theory, the action functional is quadratic in the
expansion, shear, vorticity, and acceleration. We will add to this
action a term quadratic in the spin rotation, $\kappa^2 =
\kappa_{\mu\nu} \kappa^{\mu\nu}$, and a term that couples the spin
rotation and the vorticity, $\omega\cdot\kappa = \omega_{\mu\nu}
\kappa^{\mu\nu}$. We will also include a parity violating term that
couples the spin rotation and the acceleration, $\kappa\cdot a =
\kappa_{\mu\nu} a^{\mu\nu}$. A similar term $\omega\cdot a$ that
couples the vorticity and the acceleration is a total derivative (see
Equation \eqref{eq:constr3} below). We will thus consider the
following action functional:
\begin{equation}
  \label{eq:action}
S[e_a{}^\mu] = \frac{1}{16\pi G} \int  d^4x \sqrt{-g} ( R + {\cal L}_e )
\end{equation}
where $R$ is the Ricci scalar and
\begin{equation}
{\cal L}_e = \frac{1}{3} c_\theta \theta^2 + c_\sigma
\sigma^2 + c_\omega \omega^2 +  2 c_{\omega\kappa} \omega\cdot
\kappa + c_\kappa \kappa^2  + 2 c_{\kappa a} \kappa\cdot a - c_a a^2
\end{equation}
with $\sigma^2 = \sigma_{\mu\nu} \sigma^{\mu\nu}$, $\omega^2 =
\omega_{\mu\nu} \omega^{\mu\nu}$, and $a^2 = a_{\mu\nu} a^{\mu\nu}$
and where $c_\theta$, $c_\sigma$, $c_\omega$, $c_{\omega\kappa}$,
$c_\kappa$, $c_{\kappa a}$, and $c_a$ are dimensionless coupling
constants. Further action terms involving matter and other fields can
be added to the action \eqref{eq:action} which we will not do in this
paper. Using the tensor $S_{\mu\nu\rho}$, ${\cal L}_e$ can be written
in the form
\begin{equation}
{\cal L}_e = K^{\kappa\lambda\mu\rho\nu\sigma} S_{\kappa\nu\mu} S_{\lambda\sigma\rho}
\end{equation}
where the supermetric $K^{\kappa\lambda\mu\rho\nu\sigma}$ depends
algebraically on $e_a{}^\mu$. Since we implicitly assume that
$e_a{}^\mu$ is orthonormal, the form of
$K^{\kappa\lambda\mu\rho\nu\sigma}$ is not unique. We here choose
\begin{eqnarray}
K^{\kappa\lambda\mu\rho\nu\sigma} & = & u^\kappa u^\lambda \Big( c_1
g^{\mu\rho} g^{\nu\sigma} + c_2 h^{\mu\nu} h^{\rho\sigma} + c_3
h^{\mu\sigma} h^{\rho\nu}  +  c_4 u^\mu u^\rho
g^{\nu\sigma} \Big) \nonumber \\ & + & c_{\omega\kappa} \left(
u^\kappa u^\rho h^{\sigma[\mu} h^{\nu]\lambda} + u^\lambda u^\mu
h^{\nu[\rho} h^{\sigma]\kappa} \right) - c_\kappa
u^\mu u^\rho h^{\kappa\sigma} h^{\lambda\nu} + 2 c_{\kappa a} u^\mu
u^\rho \varepsilon^{\nu\sigma [\kappa} u^{\lambda]}
\end{eqnarray}
where 
\begin{eqnarray}
c_1 & = & \frac{c_\sigma + c_\omega}{2}, \\
c_2 & = & \frac{c_\theta - c_\sigma}{3}, \\
c_3 & = & \frac{c_\sigma - c_\omega}{2}, \\
c_4 & = & 2c_a - \frac{c_\sigma + c_\omega}{2}.
\end{eqnarray}

Due to the presence of the aether velocity field $u^\mu$, the action
\eqref{eq:action} is not invariant under local Lorentz boosts. Under
local spatial rotations of the tetrad field, that is, $e^i{}_\mu
\rightarrow \Lambda^i{}_j e^j{}_\mu$ where $\Lambda^i{}_j$ is a
rotation matrix, the kinematic quantities $a_\mu$,
$\omega_{\mu\nu}$, $\sigma_{\mu\nu}$, and $\theta$ are invariant since
they do not depend on $e^i{}_\mu$. However, the spin rotation
transforms according to
\begin{equation}
  \kappa_{\mu\nu} \rightarrow \kappa_{\mu\nu}
  - e_{i\mu} e^j{}_\nu \Lambda_k{}^i \partial_u \Lambda^k{}_j.
\end{equation}
From this equation follows that if terms of the form $\kappa^2$,
$\omega\cdot\kappa$ or $\kappa\cdot a$ are present in the action
\eqref{eq:action}, there is only invariance under spatial rotations if
$\partial_u \Lambda^i{}_j = 0$. This means that the presence of the
spin rotation breaks the invariance under spatial rotations in the
direction of the aether velocity field.

%%%%%%%%%%%%%%%%%%%%%%%%%%%%%%%%%%%%%%%%%%%%%%%%%%%%%%%%%%%

\section{Field Equations}

The field equations following from the action \eqref{eq:action} are
obtained by variation of the action with respect to $e_a{}^\mu$.
After contraction with $e_{a\nu}$ and taking the symmetric and
antisymmetric part, the field equations read
\begin{equation}\label{eq:fieldeq1}
 G_{\mu\nu}  =  \frac{1}{2} g_{\mu\nu} {\cal L}_e - D_\rho \left(
J^\rho{}_{(\mu\nu)} - J_{(\mu\nu)}{}^\rho \right)
 - \left(J^{\rho\sigma}{}_{(\mu} - J_{(\mu}{}^{\sigma\rho} \right) S_{\nu)\rho\sigma}
 -  J^\rho{}_{(\mu}{}^\sigma Q_{|\rho\sigma|\nu)}
 - \frac{1}{2} N^{\alpha\beta\lambda\rho\sigma\tau}{}_{(\mu\nu)}
S_{\alpha\sigma\lambda} S_{\beta\tau\rho} ,
\end{equation}
\begin{equation}
  \label{eq:fieldeq2}
  D_\rho J_{[\mu}{}^\rho{}_{\nu]} - \left(J^{\rho\sigma}{}_{[\mu} -
   J_{[\mu}{}^{\sigma\rho} \right) S_{\nu]\rho\sigma} -
   J^\rho{}_{[\mu}{}^\sigma   Q_{|\rho\sigma|\nu]}
   + \frac{1}{2}
   N^{\alpha\beta\lambda\rho\sigma\tau}{}_{[\mu\nu]}
   S_{\alpha\sigma\lambda} S_{\beta\tau\rho} = 0 
\end{equation}
where $G_{\mu\nu}$ is the Einstein tensor,
\begin{equation}
  J^{\kappa\mu\nu} = K^{\kappa\lambda\mu\rho\nu\sigma} S_{\lambda\sigma\rho} ,
\end{equation}
and
\begin{equation}
  N^{\alpha\beta\lambda\rho\sigma\tau}{}_{\mu\nu} =
  \frac{\delta K^{\alpha\beta\lambda\rho\sigma\tau}}{\delta e_a{}^\mu} e_{a\nu}.
\end{equation}
The ``momentum'' $J^{\rho\mu\nu}$ is explicitly given by
\begin{eqnarray}
  \nonumber J^{\rho\mu\nu} = \frac{1}{3} c_\theta u^\rho h^{\mu\nu}
  \theta + c_\sigma u^\rho \sigma^{\mu\nu} + c_\omega u^\rho
  \omega^{\mu\nu} + c_{\omega\kappa} \left( u^\rho \kappa^{\mu\nu} +
  u^\mu \omega^{\nu\rho} \right) + c_\kappa u^\mu \kappa^{\nu\rho}
  \\ + c_{\kappa a} u^\mu \left( a^{\nu\rho} -2 u^\rho \kappa^\nu
  \right) + 2 c_a u^\rho u^\mu a^\nu .
\end{eqnarray}  
Equations \eqref{eq:fieldeq1} represent the Einstein field equations
with the aether stress-energy tensor on the right hand side. Equations
\eqref{eq:fieldeq2} are the field equations of the aether field. In
the case $c_\kappa = c_{\omega\kappa} = c_{\kappa a} = 0$, Equations
\eqref{eq:fieldeq1} and \eqref{eq:fieldeq2} are equivalent with the
field equations of Einstein-aether theory.

If we are working at the level of the kinematic quantities --- and
not at the level of the tetrad --- the field equations have to be
supplemented by the constraint and evolution equations for the
kinematic quantities which follow from the Ricci identity
\begin{equation}
  \label{eq:ricci}
  D_\rho D_\nu e^a{}_\mu - D_\nu D_\rho e^a{}_\mu = R^\sigma{}_{\mu\nu\rho} e^a{}_\sigma
\end{equation}
where $R_{\sigma\rho\mu\nu}$ is the Riemann curvature tensor (see
also~\cite{ellis71}). This equation can be solved for
$R_{\sigma\rho\mu\nu}$ in terms of the tensor $C_{\mu\nu\rho}$
yielding
\begin{equation}
  \label{eq:riemann}
  R_{\sigma\mu\nu\rho} = D_\rho C_{\sigma\mu\nu} - D_\nu C_{\sigma\mu\rho}
  + C_{\sigma\lambda\rho}C^\lambda{}_{\mu\nu} - C_{\sigma\lambda\nu} C^\lambda{}_{\mu\rho} .
\end{equation}
The first Bianchi identity then leads to the following 16 constraint
equations for $C_{\mu\nu\rho}$.
\begin{equation}
  \label{eq:constraints}
  \mathcal{C}^\tau{}_\sigma = \varepsilon^{\mu\nu\rho\tau}
  \left( D_\rho C_{\sigma\mu\nu} + C_{\sigma\lambda\rho} C^\lambda{}_{\mu\nu} \right) = 0.
\end{equation}
Only the projections $\mathcal{C}^\tau{}_\sigma u^\sigma$ contain
solely the kinematic quantities. These four equations are the
constraint and evolution equations for the vorticity,
\begin{align}
  \label{eq:constr3}
  D_\rho \omega^\rho & = \omega \cdot a , \\
  \label{eq:constr4}
  D_u \omega^\rho & = \frac{1}{2} D_\sigma a^{\sigma\rho} + \omega^\sigma \sigma_\sigma{}^\rho
  - \frac{2}{3} \theta \omega^\rho .
\end{align}
Furthermore, from Equation \eqref{eq:riemann}, we can compute the
time-time and time-space projections of the Ricci tensor
$R_{\mu\nu}=R^\rho{}_{\mu\rho\nu}$ which can be expressed by the
kinematic quantities,
\begin{align}
  \label{eq:constr1}
  R_{\mu\nu} u^\mu u^\nu & = - \partial_u \theta -\frac{1}{3} \theta^2
  + D_\rho a^\rho -\sigma^2 + \omega^2 , \\
  \label{eq:constr2}
  R_{\mu\nu} u^\mu h^\nu_\rho & = D_\sigma \sigma^\sigma{}_\rho +
  u_\rho \sigma^2 + \sigma_{\rho\sigma} a^\sigma - D_\sigma
  \omega^\sigma{}_\rho - u_\rho \omega^2 - \omega_{\rho\sigma}
  a^\sigma -\frac{2}{3} \partial_\rho \theta +\frac{2}{3} u_\rho
  \partial_u \theta .
\end{align}

By projecting the field equations \eqref{eq:fieldeq1} onto the aether
velocity and its orthogonal spatial directions using $u^\mu$ and
$h^\mu_\nu$, they can be split into three groups of equations
consisting of a temporal-temporal equation $[E_{00}]$ obtained by
contraction with $u^\mu u^\nu$, temporal-spatial equations $[E_{0i}]$
obtained by contraction with $u^\mu h^\nu_\rho$, and spatial-spatial
equations $[E_{ij}]$ as a result of a contraction with $h^\mu_\rho
h^\nu_\sigma$.  We first note that by taking the trace of Equation
\eqref{eq:fieldeq1}, the curvature scalar is given by
\begin{equation}
  \label{eq:rscalar}
R = c_\theta \left( \partial_u\theta + \frac{2}{3}\theta^2\right) -
c_\sigma \sigma^2 - c_\omega \omega^2 - 2 c_{\omega\kappa}
\omega\cdot\kappa - c_\kappa \kappa^2 + 2 c_{\kappa a} \left( D_\rho
\kappa^\rho - \kappa\cdot a \right) - c_a \left( 2 D_\rho a^\rho - a^2
\right) .
\end{equation}
The temporal part of Equation \eqref{eq:fieldeq1}, which contains
$G_{\mu\nu}u^\mu u^\nu = R_{\mu\nu}u^\mu u^\nu - \frac{1}{2}R$, can
then be expressed solely in terms of the kinematic quantities by
using Equations \eqref{eq:constr1} and \eqref{eq:rscalar}. The result
is
\begin{eqnarray}
  \nonumber [E_{00}] && \qquad \left( 1 + \frac{c_\theta}{2}\right)
  \left(\partial_u \theta + \frac{1}{3} \theta^2 \right) + \left( 1 -
  c_\sigma \right) \sigma^2 - \left( 1 - c_\omega + 2 c_{\omega\kappa}
  \right) \omega^2 - c_\kappa \left( \kappa^2 + 2 \omega\cdot\kappa
  \right) \\ \label{eq:e00} && - c_{\kappa a} \left[ D_\rho \left(
    \omega^\rho + \kappa^\rho \right) + \left( \omega + \kappa \right)
    \cdot a \right] - \left( 1 - c_a \right) D_\rho a^\rho = 0 .
\end{eqnarray}  
In a similar way, the temporal-spatial part of the Ricci tensor
$R_{\mu\nu} u^\mu h^\nu_\rho$ can be eliminated from $[E_{0i}]$ using
Equation \eqref{eq:constr2} resulting again in equations containing
only the kinematic quantities.  The Ricci tensor is only present in
the spatial-spatial part $[E_{ij}]$ in the form $R_{\mu\nu} h^\mu_\rho
h^\nu_\sigma$. The long equations $[E_{0i}]$ and $[E_{ij}]$ are given
in Appendix A.

The antisymmetric aether field equations \eqref{eq:fieldeq2} can be
analogously split in two groups $[A_{0i}]$ and $[A_{ij}]$ by
projections using $u^\mu$ and $h^\mu_\nu$.  The spatial-spatial
equations, which are empty in Einstein-aether theory, can be written
in the form
\begin{eqnarray}
  \nonumber
 [A_{ij}] && \qquad c_\kappa \left( D_u \kappa^\mu + \theta\kappa^\mu -
 \frac{1}{2} u^\mu \kappa \cdot a \right) + c_{\omega\kappa} \left(
 D_u \omega^\mu - \kappa^\mu{}_\rho \omega^\rho + \theta\omega^\mu -
 \frac{1}{2} u^\mu \omega\cdot a \right) \\ && + c_{\kappa a} \left( D_u
 a^\mu - \kappa^\mu{}_\rho a^\rho + \theta a^\mu - \frac{1}{2} u^\mu
 a^2 \right) = 0 .
\end{eqnarray}
These equations relate the time evolution of $\kappa_\mu$, $a_\mu$, and
$\omega_\mu$.  The long equations $[A_{0i}]$ are given in Appendix A.

Due to the constraint \eqref{eq:constr1}, only the spatial part
$R_{\mu\nu}h^{\mu\nu}$ of the curvature scalar appears in the
gravitational action besides the kinematic quantities. This can be
made more transparent by expressing $R$ by $C_{\mu\nu\rho}$ using
Equation \eqref{eq:riemann}. To this end, the spatial projection
$Q_{\mu\nu\rho}$ of $C_{\mu\nu\rho}$ can be decomposed in the
following way. Since $Q_{\mu\nu\rho}$ has only spatial components, we
can define a second order tensor $Q_{\mu\nu}$ by
\begin{equation}
  Q_{\mu\nu} = \frac{1}{2}\varepsilon_{\mu\rho\sigma} Q^{\rho\sigma}{}_\nu .
\end{equation}
This tensor can be irreducibly decomposed into its antisymmetric part
$\Omega_{\mu\nu}$, its trace $\Theta$ which is the totally
antisymmetric part of $Q_{\mu\nu\rho}$, and its symmetric trace-free
part $\Sigma_{\mu\nu}$:
\begin{align}
  \Omega_{\mu\nu} & = Q_{[\mu\nu]} = \frac{1}{2}
  \varepsilon_{\rho\sigma[\mu}
    Q^{\rho\sigma}{}_{\nu]} ,\\
  \Theta & = Q^\rho{}_\rho = \frac{1}{2}
  \varepsilon_{\mu\nu\rho}
  Q^{\mu\nu\rho} ,\\
  \Sigma_{\mu\nu} & = Q_{(\mu\nu)} - \frac{1}{3}
  h_{\mu\nu}\Theta = \frac{1}{2} \varepsilon_{\rho\sigma(\mu}
  Q^{\rho\sigma}{}_{\nu)} -\frac{1}{3} h_{\mu\nu}\Theta .
 \end{align}
Defining the vector
\begin{equation}
  \Omega_\mu = \frac{1}{2} \varepsilon_{\mu\rho\sigma}\Omega^{\rho\sigma}
  = \frac{1}{2} Q_{\mu\rho}{}^\rho ,
\end{equation}
the irreducible decomposition of $Q_{\mu\nu\rho}$ reads
\begin{equation}
  \label{eq:q}
  Q_{\mu\nu\rho} = 2 \Omega_{[\mu} h_{\nu]\rho}
  - \varepsilon_{\mu\nu\sigma} \Sigma^\sigma{}_\rho
  - \frac{1}{3} \varepsilon_{\mu\nu\rho} \Theta .
\end{equation}
Using Equation \eqref{eq:riemann}, the Ricci scalar is then given by
\begin{equation}
  R = - \frac{2}{3} \Theta^2 + \Sigma^2 - \Omega^2 + 2 \Omega\cdot a +
  \frac{2}{3} \theta^2 - \sigma^2 + \omega^2 + 2 \omega\cdot\kappa - 2
  D_\rho \left( 2 \Omega^\rho + u^\rho \theta - a^\rho \right)
\end{equation}
where $\Sigma^2 = \Sigma_{\mu\nu}\Sigma^{\mu\nu}$, $\Omega^2 =
\Omega_{\mu\nu}\Omega^{\mu\nu}$, and $\Omega\cdot a = \Omega_{\mu\nu}
a^{\mu\nu}$.  Inserting this equation into the Lagrangian ${\cal L} =
R + {\cal L}_e$, we obtain
\begin{equation}
  {\cal L} = - \frac{2}{3} \Theta^2 + \Sigma^2 - \Omega^2 + 2  \Omega\cdot a
  + \frac{2}{3} \left( 1 + \frac{c_\theta}{2} \right) \theta^2
  - \left( 1 - c_\sigma \right) \sigma^2
  + \left( 1 + c_\omega \right) \omega^2
  +  2 \left( 1 + c_{\omega\kappa} \right) \omega\cdot\kappa
  + c_\kappa \kappa^2  + 2 c_{\kappa a} \kappa\cdot a - c_a a^2
\end{equation}
where total derivatives have been omitted. While the quantities
$\Omega_\mu$, $\Sigma_{\mu\nu}$, and $\Theta$ appear together with the
kinematic quantities in the decomposition of $C_{\mu\nu\rho}$, they
have a different status than the kinematic quantities since they
depend on the arbitrary orientation of the triad in space. The
kinematic quantities can be seen as field strengths with the tetrad
as potentials.

Note that while the quantities $\Omega_\mu$, $\Sigma_{\mu\nu}$, and
$\Theta$ are not invariant under space dependent spatial rotations,
the special combination $-\frac{2}{3} \Theta^2 + \Sigma^2 - \Omega^2 +
2 \Omega\cdot a$ is invariant under such transformations up to a total
derivative. By adding general terms quadratic in $\Omega_\mu$,
$\Sigma_{\mu\nu}$, and $\Theta$ as well as cross terms with the
kinematic quantities to the Lagrangian, the invariance under the full
Lorentz group could be broken which we will not do in this paper.

%%%%%%%%%%%%%%%%%%%%%%%%%%%%%%%%%%%%%%%%%%%%%%%%%%%

\section{Weak Fields}

In this section, we will study the linearized version of the field
equations \eqref{eq:fieldeq1} and \eqref{eq:fieldeq2}. In the weak
field approximation, the vectors $e_a{}^{\mu}$ and $e^a{}_{\mu}$ can
be expanded around Minkowski space-time up to first order terms
according to
\begin{eqnarray}
e_a{}^{\mu} & = & \delta^\mu_a + \chi_a{}^{\mu} ,\\
e^a{}_{\mu} & = & \delta^a_\mu + \psi^a{}_\mu
\end{eqnarray}
where indices of $\chi_a{}^{\mu}$ and $\psi^a{}_\mu$ are raised and
lowered with the Minkowski metric. In the following, we will therefore
not differentiate between internal and spacetime indices of first
order fields. The condition $\delta^\mu_\nu = e_a{}^{\mu} e^a{}_{\nu}$
leads to the relation $\chi_{\mu\nu} = -\psi_{\nu\mu}$ which allows to
use only $\psi_{\mu\nu}$ in the following. $\psi_{\mu\nu}$ can be
decomposed into its symmetric and antisymmetric part according to
\begin{equation}
  \label{eq:psidecomp}
\psi_{\mu\nu} = \frac{1}{2} \gamma_{\mu\nu} + \zeta_{\mu\nu}
\end{equation}
where $\gamma_{\mu\nu} = \psi_{\mu\nu} + \psi_{\nu\mu}$ and $\zeta_{\mu\nu} =
\psi_{[\mu\nu]}$. The metric tensor is then given up to first order by
\begin{equation}
g_{\mu\nu} = \eta_{\mu\nu} + \gamma_{\mu\nu} .
\end{equation}

Under infinitesimal coordinate transformations $x^\mu \to x^\mu +
\xi^\mu$, the tetrad transforms as
\begin{equation}
  \delta e_a{}^\mu = {\cal L}_\xi e_a{}^\mu = \xi^\nu D_\nu
  e_a{}^\mu - e_a{}^\nu D_\nu \xi^\mu = -\partial_a \xi^\mu .
\end{equation}
This leads to the following gauge transformations of $\gamma_{\mu\nu}$
and $\zeta_{\mu\nu}$.
\begin{eqnarray}
  \label{eq:gaugetr1}
  \gamma_{\mu\nu} & \to & \gamma_{\mu\nu} + \partial_\mu \xi_\nu
  + \partial_\nu \xi_\mu , \\
  \label{eq:gaugetr2}
  \zeta_{\mu\nu} & \to & \zeta_{\mu\nu} - \partial_{[\mu} \xi_{\nu]} .
\end{eqnarray}

In order to simplify the notation, in the following all fields will be
first order quantities in this section. The tensor $C_{\mu\nu\rho}$ in
first order approximation is
\begin{equation}
  \label{eq:c_lin}
  C_{\mu\nu\rho} = \partial_\rho \zeta_{\mu\nu} + \partial_{[\mu} \gamma_{\nu]\rho} .
\end{equation}
It can be checked that $C_{\mu\nu\rho}$ is gauge invariant implying
that all kinematic quantities in first order approximation are gauge
invariant.

The field equations \eqref{eq:fieldeq1} and \eqref{eq:fieldeq2} in
first order are given by
\begin{align}
{G}_{\mu\nu} = \partial_\rho
{J}_{(\mu\nu)}{}^\rho - \partial_\rho
{J}{}^\rho{}_{(\mu\nu)} ,\\
\partial_\rho {J}{}_{[\mu}{}^\rho{}_{\nu]} = 0 .
\end{align}
Since the first order approximation of the kinematic quantities as
well as the first order curvature tensor are gauge invariant, it is
convenient to write the first order field equations in terms of the
kinematic quantities. The first order approximations of the aether
field equations $[A_{ij}]$ and $[A_{0i}]$ are
\begin{align}
\label{eq:aij}
        [A_{ij}] & \qquad c_{\omega\kappa} \bm{\dot{\omega}} + c_\kappa
        \bm{\dot{\kappa}} + c_{\kappa a} \bm{\dot{a}} = \bm{0} , \\
  \label{eq:a0i}
        [A_{0i}] &\qquad \frac{1}{3}c_\theta \bm{\nabla} \theta - c_\sigma
        \bm{\nabla\cdot\sigma} - c_\omega \bm{\nabla\times\omega}
        -c_{\omega\kappa} \bm{\nabla\times\kappa} -
        2 c_{\kappa a}  \bm{\dot{\kappa}}
        + 2 c_a  \bm{\dot{a}} = \bm{0} .
\end{align}
Here and in the following, we use vector notation. For example
$(\bm{a})_i = a_i$, $(\bm{\nabla\times \omega})_i =
\epsilon_{ijk}\partial_j\omega_k = \partial_j \omega_{ji}$,
$\bm{\nabla\cdot \kappa}=\partial_i\kappa_i$, $\dot{\theta} =
\partial_0 \theta$, $(\bm{\sigma})_{ij} = \sigma_{ij}$, $(\bm{R})_{ij}
= R_{ij}$, $(\bm{1})_{ij} = \delta_{ij}$. ($\epsilon_{ijk}$ is the
totally antisymmetric symbol with $\epsilon_{123} = 1$.)  The first
order approximations of the field equations $[E_{00}]$ and $[E_{0i}]$
lead to
\begin{align}
 \label{eq:0i}
       [I_0] &  \qquad c_{\kappa a} \bm{\nabla\cdot \kappa}
       + \left( 1 - c_a \right) \bm{\nabla\cdot a} =
         -\left( 1 + \frac{c_\theta}{2} \right) \dot{\theta} , \\
   \nonumber
         [I_i] & \qquad \left( c_\omega + \left( 1 + c_{\omega\kappa} \right)
         \frac{c_\sigma}{1-c_\sigma}
         \right) \bm{\nabla\times \omega} +
         \left( c_{\omega\kappa} +  c_\kappa \frac{c_\sigma}{1-c_\sigma}
         \right) \bm{\nabla\times \kappa}\\
   \label{eq:ii}
         & \qquad + 2c_{\kappa a}  \bm{\dot{\kappa}}
   + c_{\kappa a} \frac{c_\sigma}{1-c_\sigma}  \bm{\nabla\times a}
   - 2 c_a \bm{\dot{a}}
         = \frac{2c_\sigma + c_\theta}{3\left( 1-c_\sigma \right)}
         \bm{\nabla} \theta
\end{align}
where in Equation \eqref{eq:ii} the field equation $[A_{0i}]$ has been
substituted in order to eliminate $\bm{\nabla\cdot\sigma}$.
Finally, the weak field version of the field equation $[E_{ij}]$ is
\begin{align}
\label{eq:iij}
         [I_{ij}] & \qquad \bm{R} = - \frac{1}{6} c_\theta \bm{1} \dot{\theta}
         - c_\sigma \bm{\dot{\sigma}}
         + c_{\kappa a} \bm{1}  \bm{\nabla\cdot \kappa}
         - c_a \bm{1} \bm{\nabla\cdot a} .
\end{align}

Since we are working at the level of the kinematic quantities, we have
to take the corresponding constraint equations into account.  The
linear approximations of the constraint equations \eqref{eq:constr3}
and \eqref{eq:constr4} are
\begin{align}
 \label{eq:h0}
        [H_0]  & \qquad \bm{\nabla\cdot\omega} = 0 ,\\
   \label{eq:hi}
        [H_i]  & \qquad \bm{\nabla\times a} + 2 \bm{\dot{\omega}}= \bm{0} .
\end{align}

The spin rotation $\bm{\kappa}$ can be eliminated from the weak field
equations using Equation \eqref{eq:aij}. For this, we can split the
fields into time independent and time varying parts. The static part
corresponds to the zero frequency Fourier mode of the fields while the
dynamic part represents the finite frequency contribution.  We here
consider only the dynamic case. Equation \eqref{eq:aij} can then be
integrated resulting in
\begin{equation}\label{eq:kappa} 
  \bm{\kappa} = -\frac{c_{\omega\kappa}}{c_\kappa} \bm{\omega}
    - \frac{c_{\kappa a}}{c_\kappa} \bm{a} .
\end{equation}
Substituting this equation into the field equations $[I_0]$, $[I_i]$, and
$[I_{ij}]$, we obtain
\begin{eqnarray}
  \label{eq:dyn_i0}
  &[I_0] & \qquad 
  \bm{\nabla\cdot a} = -\frac{ 1 + \frac{c_\theta}{2}}{1 - \bar{c}_a}
   \dot{\theta} ,\\
  \label{eq:dyn_ii}
  &[I_i] & \qquad \frac{1}{2 \bar{c}_a} \left( \bar{c}_\omega
  + \frac{c_{\sigma}}{1-c_\sigma}
         \right) \bm{\nabla\times \omega}
         - \bm{\dot{a}}
         = \frac{c_\sigma + \frac{c_\theta}{2}}{3\bar{c}_a\left( 1-c_\sigma \right)}
         \bm{\nabla} \theta ,\\
  \label{eq:dyn_iij}
  &[I_{ij}] & \qquad \bm{R} = -\frac{1}{3}\left[ \frac{c_\theta}{2}
    - \frac{3\bar{c}_a\left( 1 +  \frac{c_\theta}{2} \right)}{1-\bar{c}_a}
    \right] \bm{1} \dot{\theta}
         - c_\sigma \bm{\dot{\sigma}}       
\end{eqnarray}
where
\begin{eqnarray}
  \label{eq:sub1}
 \bar{c}_\omega = c_\omega - \frac{c_{\omega\kappa}^2}{c_\kappa} ,\\
  \label{eq:sub2}
  \bar{c}_a = c_a + \frac{c_{\kappa a}^2}{c_\kappa} .
\end{eqnarray}
Furthermore, combining Equations \eqref{eq:a0i} and \eqref{eq:dyn_ii}
yields
\begin{equation}
  \label{eq:dyn_a0i}
        [A_{0i}] \qquad \bm{\nabla\times \omega} =
        \frac{2}{3}\left( 1 + \frac{c_\theta}{2} \right)
        \bm{\nabla}\theta
        + \left( 1-c_\sigma \right) \bm{\nabla\cdot\sigma} .
\end{equation}
Comparing Equations \eqref{eq:dyn_i0}-\eqref{eq:dyn_iij} with the
limiting case $c_\kappa = c_{\omega\kappa} = c_{\kappa a} = 0$ of
Equations \eqref{eq:0i}-\eqref{eq:iij}, that is, Einstein-aether
theory, we conclude that the introduction of the spin rotation amounts
in the dynamic case to the substitutions $c_\omega \rightarrow
\bar{c}_\omega$ and $c_a \rightarrow \bar{c}_a$.

Equations \eqref{eq:h0}, \eqref{eq:hi}, \eqref{eq:dyn_i0}, and
\eqref{eq:dyn_ii} have a close resemblance with Maxwell's equations
where $\bm{a}$ and $2\bm{\omega}$ play
the role of electric and magnetic fields, respectively, and where the
electric charge density $\rho$ and current $\bm{j}$ are given by
 \begin{align}
   \label{eq:rho}
   \rho &= - \frac{1+\frac{c_\theta}{2}}{1-\bar{c}_a}\dot{\theta} ,\\
   \label{eq:j}
  \bm{j} &= \frac{c_\sigma +\frac{c_\theta}{2}}{3\bar{c}_a\left(1-c_\sigma\right)}
  \bm{\nabla}\theta .
\end{align}

A further useful equation involving the kinematic quantities follows
from the first order limit of the contracted second Bianchi identity
and the field equation $[I_{ij}]$ (see Appendix B for a derivation):
\begin{equation}
  \label{eq:bianchi}
\Delta\bm{\sigma} - \left( 1 - c_\sigma \right) \bm{\ddot{\sigma}}
- 2 \left(\nabla\left(\nabla\cdot\bm{\sigma}\right)\right)_{\textrm{sym}}
 - \frac{1}{3}\bm{1} \Delta \theta
+ \frac{1}{3} \left( 1 + \frac{c_\theta}{2} \right)
 \left( 1 - \frac{3\bar{c}_a}{1-\bar{c}_a} \right)
  \bm{1} \ddot{\theta} - \frac{1}{3} \nabla\nabla\theta +
 \left( \nabla\dot{\bm{a}}\right)_{\textrm{sym}} = \bm{0}
\end{equation}
where the suffix sym denotes symmetrization.
  
The linearized wave solutions of Einstein-aether theory were given
in~\cite{jacobson04}.  Since the weak field equations of the extended
theory are effectively the same as in Einstein-aether theory, the wave
solutions are the same but with different wave speeds and field
content.  The analogy of the weak field equations with Maxwell's
equations suggests the existence of spin 1 acceleration-vorticity
waves similar to electromagnetic waves. Indeed, in the case $\theta=0$
Equations \eqref{eq:hi} and \eqref{eq:dyn_ii} lead to the wave
equations
\begin{equation}\label{eq:wave_aomega}
  \ddot{\bm{a}} - s_{a\omega}^2 \Delta{\bm{a}} = \bm{0},\qquad
  \ddot{\bm{\omega}} - s_{a\omega}^2 \Delta{\bm{\omega}} = \bm{0}
\end{equation}
with the wave speed $s_{a\omega}$ given by
\begin{equation}
  s_{a\omega}^2 = \frac{1}{4\bar{c}_a} \left( \bar{c}_\omega
    + \frac{c_{\sigma}}{1-c_\sigma} \right) .
\end{equation}
From Equations \eqref{eq:h0}, \eqref{eq:dyn_i0}, and \eqref{eq:hi}
follows that $\bm{a}$ and $\bm{\omega}$ are transverse and
perpendicular to each other.  Equation \eqref{eq:kappa} with Equations
\eqref{eq:wave_aomega} leads to a wave equation for $\bm{\kappa}$ with
a polarization determined by the coupling constants
$c_{\omega\kappa}$, $c_{\kappa a}$, and $c_\kappa$.

Furthermore, the conservation equation $\dot{\rho} +
\bm{\nabla}\cdot\bm{j} = 0$ for the charge density \eqref{eq:rho} and
the current \eqref{eq:j} corresponds to a wave equation for spin 0
expansion waves,
\begin{equation}
  \ddot{\theta} - s_\theta^2 \Delta \theta = 0
\end{equation}
where the wave speed $s_\theta$ is given by
\begin{equation}
  s_\theta^2 = \frac{1 - \bar{c}_a}{3\bar{c}_a}
  \left( \frac{\frac{c_\theta}{2}}{1+\frac{c_\theta}{2}} + \frac{c_\sigma}{1-c_\sigma} \right).
\end{equation}
From Equation \eqref{eq:dyn_i0} follows that these waves are
accompanied by longitudinal acceleration waves.  Equation
\eqref{eq:hi} then shows that for these solutions
$\bm{\omega}=\bm{0}$.  From Equation \eqref{eq:kappa} follows that the
spin rotation waves are also longitudinal.

Finally, in the case $\theta=0$,
$\bm{a}=\bm{\omega}=\bm{\kappa}=\bm{0}$ it follows from Equation
\eqref{eq:dyn_a0i} that $\nabla\cdot\bm{\sigma}=\bm{0}$ and Equation
\eqref{eq:bianchi} leads to a wave equation for spin 2 transverse
shear waves,
\begin{equation}
  \ddot{\bm{\sigma}} - s_\sigma^2 \Delta\bm{\sigma} = \bm{0}
\end{equation}
with the wave speed squared
\begin{equation}
  s_\sigma^2 = \frac{1}{1-c_\sigma}.
\end{equation}
The plane wave solutions for the three cases are given in Appendix C.

If the weak field equations are expressed by the tetrad variables, we
have to take the gauge transformations \eqref{eq:gaugetr1} and
\eqref{eq:gaugetr2} into account. In order to identify the independent
wave modes, a gauge fixing has to be applied. In the present case, we
use the gauge fixing conditions
\begin{eqnarray}
  \psi_{i0} = 0 , \\
  \partial_i \psi_{0i} = 0 .
\end{eqnarray}
Given  a field configuration $\psi_{\mu\nu}$, these conditions can be reached
by applying the gauge transformation
\begin{align}
  \xi_0 & = \int \frac{d^3x'}{4\pi |x-x'|} \partial_i \psi_{0i} ,\\
  \xi_i & = - \int dt' \psi_{i0} .
\end{align}
In terms of the decomposition \eqref{eq:psidecomp}, the gauge fixing reads
\begin{eqnarray}
  \zeta_{i0} = - \frac{1}{2} \gamma_{i0} , \\
  \partial_i \zeta_{0i} = 0, \quad \partial_i \gamma_{0i} = 0 .
\end{eqnarray}
In the following, we will again use vector notation to simplify
equations.  We define $(\bm A)_i = 2\zeta_{i0} = -\gamma_{i0}$,
$(\bm\zeta)_i = - \frac{1}{2}\epsilon_{ijk}\zeta_{jk}$, $\gamma =
\frac{1}{2} \gamma_{kk}$, $(\bm\gamma)_{ij} = \frac{1}{2} \gamma_{ij}
- \frac{1}{3}\delta_{ij}\gamma$, $\phi = \frac{1}{2}
\gamma_{00}$. The gauge fixing conditions can then be written as a
kind of Coulomb gauge,
\begin{equation}
  \label{eq:gauge}
  \nabla\cdot\bm A = 0 .
\end{equation}
Taking the gauge fixing into account, the kinematic quantities in terms of
the tetrad variables are
\begin{align}
  \bm{a} & = - \dot{\bm A} - \nabla \phi ,\\
  \bm{\omega} & = \frac{1}{2} \nabla \times \bm{A} ,\\
  \bm{\kappa} & = - \dot{\bm{\zeta}} - \frac{1}{2} \nabla\times \bm{A} ,\\
  \bm{\sigma} & = \dot{\bm{\gamma}} ,\\
  \theta & = - \dot{\gamma} .
\end{align}
The tetrads for plane waves are given in Appendix C.

%%%%%%%%%%%%%%%%%%%%%%%%%%%%%%%%%%%%%%%%%%%%%%%%%%%%%%%%%%%%%%%%%%%%%%%%%

\section{Geometrical Considerations}

In this section, we resume the discussion of the nonlinear theory of
the spinning aether in sections II and III.  The formulation given
there is entirely within Riemannian geometry, that is, gravity is
described by the metric part of the tetrad. The aether, in contrast,
is described by the full tetrad. Moreover, the symmetries of the
aether --- the broken Lorentz invariance --- are imposed at the
dynamical level through the action functional. In the following, we
will argue for a pure geometrical formulation of the aether which
implements the gravitational nature of the aether and its symmetries
already at the kinematical level.

For a start, we observe that we can define a covariant derivative
$\stackrel{*}{D}_\mu\!$ by
\begin{equation}
  \label{eq:dstar}
  \stackrel{*}{D}_\mu\!\! X^\nu = D_\mu X^\nu + S^\nu{}_{\rho\mu} X^\rho
\end{equation}
where $X^\mu$ is a vector field and $S_{\mu\nu\rho}$ is given by
Equation \eqref{eq:stensor}. Since $S_{\mu\nu\rho}$ is antisymmetric
in the first two indices, the connection corresponding to
$\stackrel{*}{D}_\mu\!$ is metric compatible, that is,
$\stackrel{*}{D}_\mu\!\! g_{\nu\rho}=0$.  With the help of Equations
\eqref{eq:irrdec}, \eqref{eq:spinrot2}, and \eqref{eq:stensor}, we can
show that for a preferred frame $e_a{}^\mu$
\begin{align}
  \label{eq:dstar2}
  \stackrel{*}{D}_\mu \!\! u^\nu = 0,\\
  \stackrel{*}{D}_u \!\! e_i{}^\mu = 0.
\end{align}
These equations may be viewed as trivial rearrangements of Equations
\eqref{eq:irrdec} and \eqref{eq:spinrot2}. However, they can also be
interpreted as defining a non-Riemannian geometry in which the aether
velocity $u^\mu$ is a parallel vector field and the triad $e_i{}^\mu$
is parallel in the direction of $u^\mu$. In this geometry,
$S_{\mu\nu\rho}$ is the contortion tensor with the torsion tensor
\begin{equation}
  T_{\mu\nu\rho} = S_{\mu\rho\nu} - S_{\mu\nu\rho}.
\end{equation}
Thus, in this geometry, the kinematic quantities are torsion fields
which realizes mathematically the interpretation of the kinematic
quantities as field strengths.

In order to gain insight into the geometry of $\stackrel{*}{D}_\mu\!$,
we can compute the corresponding spin connection. The spin connection
$\stackrel{*}{\omega}{\!}^a{}_{b\mu}$ is related to the connection
coefficients $\stackrel{*}{\Gamma}{\!}^\mu{}_{\nu\rho}$ of the
derivative $\stackrel{*}{D}_\mu$ by
\begin{equation}
  \stackrel{*}{\omega}{\!}^a{}_{b\mu} = e^a{}_\nu \partial_\mu
  e_b{}^\nu + e^a{}_\nu \stackrel{*}{\Gamma}{\!}^\nu{}_{\lambda\mu}
  e_b{}^\lambda =  e^a{}_\nu \stackrel{*}{D}_\mu\! e_b{}^\nu.
\end{equation}
This equation corresponds to a local linear transformation of the
connection coefficients from a holonomic (coordinate) basis to an
anholonomic basis given by the tetrad. In the case of the derivative
\eqref{eq:dstar}, it can be shown that
\begin{align}
  \label{eq:spinconn1}
    & \stackrel{*}{\omega}{\!}^i{}_{0\mu} = 0 ,\\
    \label{eq:spinconn2}
    &  \stackrel{*}{\omega}{\!}^i{}_{j\rho} u^\rho = 0 ,\\
    \label{eq:spinconn3}
    &  \stackrel{*}{\omega}{\!}^i{}_{j\rho} h^\rho_\mu = \omega^i{}_{j\rho} h^\rho_\mu .
\end{align}
where $\omega^a{}_{b\mu}$ is the torsion-free Levi-Civita
connection. Equations \eqref{eq:spinconn1}-\eqref{eq:spinconn3} can be
summarized as
\begin{equation}
  \stackrel{*}{\omega}{\!}^a{}_{b\mu} = - Q^a{}_{b\mu}
\end{equation}
where $Q^a{}_{b\mu} = e^a{}_\rho e_b{}^\sigma Q^\rho{}_{\sigma\mu}$.
Equation \eqref{eq:spinconn1} means that the connection
$\stackrel{*}{\omega}{\!\!}^a{}_{b\mu}$ is a $SO(3)$ connection.
Equation \eqref{eq:spinconn2} means additionally that
$\stackrel{*}{\omega}{\!}^a{}_{b\mu}$ is a trivial $\mathbf{1}$
connection along $u^\mu$. According to Equation \eqref{eq:spinconn3},
the spatial geometry is Riemannian. The connection
\eqref{eq:spinconn1} was used in~\cite{gasperini87} to formulate a
version of Einstein-aether theory employing Weinberg's
quasi-Riemannian gravity. The full connection
\eqref{eq:spinconn1}-\eqref{eq:spinconn3} was derived from the
symmetries of the spinning aether in~\cite{kohler00}. The
corresponding geometry was referred to as semi-teleparallel since only
the temporal part of space-time is parallelized.

From Equation \eqref{eq:spinconn1} follows that the time-space
components of the curvature tensor
$R^{ab}{}_{\mu\nu}=2\partial_{[\mu}\omega^{ab}{}_{\nu]} +
2\omega^a{}_{c[\mu} \omega^{cb}{}_{\nu]}$ vanish,
\begin{equation}
  \label{eq:r0i}
 \stackrel{*}{R}{\!}^{i0}{}_{\rho\sigma} = 2\partial_{[\rho}
   \stackrel{*}{\omega}{\!}^{i0}{}_{\sigma]} + 2
 \stackrel{*}{\omega}{\!}^i{}_{j[\rho}  \stackrel{*}{\omega}{\!}^{j0}{}_{\sigma]} = 0 .
\end{equation}
Using space-time indices, Equation \eqref{eq:r0i} means
\begin{equation}
  \label{eq:r0nu}
  u_\mu\!  \stackrel{*}{R}{\!}^{\mu\nu}{}_{\rho\sigma} = 0 .
\end{equation}
From the Ricci identity for the derivative \eqref{eq:dstar},
\begin{equation}
  \label{eq:riccistar}
  \stackrel{*}D_\rho \stackrel{*}D_\nu \! e^a{}_\mu - \stackrel{*}D_\nu
  \stackrel{*}D_\rho \! e^a{}_\mu
  + T^\sigma{}_{\rho\nu}\! \stackrel{*}D_\sigma \! e^a{}_\mu =
  \stackrel{*}R{\!}^\sigma{}_{\mu\nu\rho} e^a{}_\sigma ,
\end{equation}  
follows a relation between the curvature tensors of the derivatives
$\stackrel{*}D_\mu$ and $D_\mu$,
\begin{equation}\label{eq:rstar}
  \stackrel{*}R{\!}_{\sigma\mu\nu\rho} = R_{\sigma\mu\nu\rho} - D_\rho
  S_{\sigma\mu\nu} + D_\nu S_{\sigma\mu\rho} - S_{\sigma\lambda\rho}
  S^\lambda{}_{\mu\nu} + S_{\sigma\lambda\nu} S^\lambda{}_{\mu\rho}.
\end{equation}
Using Equations \eqref{eq:r0nu} and \eqref{eq:rstar}, the relations
\eqref{eq:constr1} and \eqref{eq:constr2} can be derived which can be
seen to be a direct consequence of the semi-teleparallel geometry.
Moreover, the first Bianchi identity for the semi-teleparallel
connection is equivalent with the constraint equations
\eqref{eq:constraints}.

The action functional \eqref{eq:action} can be directly interpreted in
terms of the semi-teleparallel geometry. A more general action, which
is not equivalent with \eqref{eq:action}, can be obtained if we relax
the implicit assumption that the spatial geometry is Riemannian, that
is, has vanishing torsion. In this case, however, we go beyond the
formulation of Einstein-aether theory within Riemannian geometry in
that we start with a Riemann-Cartan geometry. In this geometry, we use
a general Lorentz connection
\begin{equation}
  \tilde{\omega}^a{}_{b\mu} = \omega^a{}_{b\mu} + \tilde{K}^a{}_{b\mu}
\end{equation}
where $\tilde{K}^a{}_{b\mu} = e^a{}_\rho e_b{}^\nu
\tilde{K}^\rho{}_{\nu\mu}$ is the contortion tensor. In order to
formulate a theory of gravity with a spinning aether using
Riemann-Cartan geometry, the action \eqref{eq:action} has to be
extended to an Einstein-Cartan theory.  The natural way is to define
the kinematic quantities with respect to the connection
$\tilde{\omega}^a{}_{b\mu}$ and to use the curvature scalar belonging
to $\tilde{\omega}^a{}_{b\mu}$ in the Lagrangian. The kinematic
quantities corresponding to the derivative $\tilde{D}_\mu$ are (see
also \cite{capozziello01})
\begin{align}
  \label{eq:atilde}
  \tilde{a}_\mu & = a_\mu + \tilde{K}_{\mu\rho\sigma} u^\rho u^\sigma , \\
  \label{eq:omegatilde}
  \tilde{\omega}_{\mu\nu} &= \omega_{\mu\nu} +  \tilde{K}^\rho{}_{[\mu\nu]} u_\rho
  - u_{[\mu}  \tilde{K}_{\nu]\rho\sigma} u^\rho u^\sigma ,\\
  \label{eq:thetatilde}
  \tilde{\theta} & = \theta - \tilde{K}_\rho u^\rho , \\
  \label{eq:sigmatilde}
  \tilde{\sigma}_{\mu\nu} & = \sigma_{\mu\nu} - \tilde{K}^\rho{}_{(\mu\nu)} u_\rho
  +  u_{(\mu}  \tilde{K}_{\nu)\rho\sigma} u^\rho u^\sigma + \frac{1}{3} h_{\mu\nu}
  \tilde{K}_\rho u^\rho , \\
  \label{eq:kappatilde}
  \tilde{\kappa}_{\mu\nu} & = \kappa_{\mu\nu} +  \tilde{K}_{\mu\nu\rho} u^\rho
  + 2 u_{[\mu}  \tilde{K}_{\nu]\rho\sigma} u^\rho u^\sigma
\end{align}
where $ \tilde{K}^\mu = \tilde{K}^{\mu\rho}{}_\rho $ is the trace of
the contortion tensor.  Furthermore, the curvature scalar
corresponding to $\tilde{D}_\mu$ is
\begin{equation}
  \tilde{R} = R + 2 D_\rho \tilde{K}^\rho - \tilde{K}_\rho \tilde{K}^\rho
  + \tilde{K}_{\sigma\lambda\rho} \tilde{K}^{\rho\lambda\sigma}.
\end{equation}
The generalization of the action \eqref{eq:action} to an
Einstein-Cartan-aether theory thus is
\begin{eqnarray}
  \label{eq:ec_action}
  \nonumber
  S[e_a{}^\mu,\tilde{\omega}^a{}_{b\mu}] & = &\frac{1}{16\pi G} \int d^4x
  \sqrt{-g} \big( R - \tilde{K}_\rho \tilde{K}^\rho +
  \tilde{K}_{\sigma\lambda\rho} \tilde{K}^{\rho\lambda\sigma} \\
  & + & \frac{1}{3} \tilde{c}_\theta \tilde{\theta}^2 + \tilde{c}_\sigma
  \tilde{\sigma}^2 + \tilde{c}_\omega \tilde{\omega}^2 + 2
  \tilde{c}_{\omega\kappa} \tilde{\omega}\cdot \tilde{\kappa} +
  \tilde{c}_\kappa \tilde{\kappa}^2 + 2 \tilde{c}_{\kappa a}
  \tilde{\kappa}\cdot \tilde{a} - \tilde{c}_a \tilde{a}^2 \big)
\end{eqnarray}
up to a surface term.

In order to get an action functional for a semi-teleparallel geometry,
simply replacing the Lorentz connection in the action for
Einstein-Cartan theory \eqref{eq:ec_action} by a semi-teleparallel
connection is not suitable since the corresponding kinematic
quantities \eqref{eq:atilde}-\eqref{eq:kappatilde} are zero in this
case and the scalar curvature contains only the spatial part of the
curvature tensor.  A method to obtain an action functional for a
semi-teleparallel geometry was proposed in~\cite{kohler00}. It was
shown there that starting from a general Lorentz connection
$\tilde{\omega}^a{}_{b\mu}$, there is a unique decomposition
\begin{equation}
  \label{eq:omegadec}
  \tilde{\omega}^a{}_{b\mu} =\; \stackrel{*}{\omega}{\!}^a{}_{b\mu} +
  H^a{}_{b\mu}
\end{equation}
where $\stackrel{*}{\omega}{\!}^a{}_{b\mu}$ is a semi-teleparallel
connection with respect to a tetrad $e_a{}^\mu$ and where
$H^a{}_{b\mu}$ is a tensor field satisfying $H_{ab\mu}=-H_{ba\mu}$ and
which has vanishing spatial components, $H_{ij\mu}h^\mu_\nu =
0$. Actually, $H^a{}_{b\mu}$ is the difference between the contortion
tensors of $\tilde{\omega}^a{}_{b\mu}$ and
$\stackrel{*}{\omega}{\!}^a{}_{b\mu}$,
\begin{equation}
  H^\mu{}_{\nu\rho} =\;
  \tilde{K}^\mu{}_{\nu\rho} - \stackrel{*}{K}{\!}^\mu{}_{\nu\rho}.
\end{equation}
The basic idea in constructing an action for the semi-teleparallel
geometry is the following:
\begin{enumerate}
  \item Start with an action functional
    $S[e_a{}^\mu,\tilde{\omega}^a{}_{b\mu}]$ that depends on the
    tetrad and an arbitrary Lorentz connection
    $\tilde{\omega}^a{}_{b\mu}$.
  \item Insert the decomposition \eqref{eq:omegadec} to obtain an
    action functional
    $S[e_a{}^\mu,\stackrel{*}{\omega}{\!}^a{}_{b\mu},H^a{}_{b\mu}]$.
  \item Find a stationary point of this action with respect to
    $H^a{}_{b\mu}$ resulting in an action functional
    ${S'}[e_a{}^\mu,\stackrel{*}{\omega}{\!}^a{}_{b\mu}]$ that is
    defined for a semi-teleparallel connection
    $\stackrel{*}{\omega}{\!}^a{}_{b\mu}$.
\end{enumerate}
In the present case, we start with the action \eqref{eq:ec_action}.
Insertion of the decomposition \eqref{eq:omegadec} and variation with
respect to $H_{\mu\nu\rho}$ leads to algebraic equations for
$H_{\mu\nu\rho}$. Solving these equations and inserting
$H_{\mu\nu\rho}$ back into the action \eqref{eq:ec_action} results in
a new action of the form
\begin{eqnarray}
  \label{eq:st_action}
  \nonumber
  {S'}[e_a{}^\mu,\stackrel{*}{\omega}{\!}^a{}_{b\mu}] & = &
  \frac{1}{16\pi G} \int d^4x \bigg[ R + \frac{1}{3} c_\theta \theta^2
  + c_\sigma \sigma^2 + c_\omega \omega^2 + 2 c_{\omega\kappa}
  \omega\cdot \kappa + c_\kappa \kappa^2  \\
  & + & \left( 2 a_\rho  + c_{\omega Z} \omega_\rho  +
  c_{\kappa Z} \kappa_\rho \right) Z^\rho + c_{Z} Z_\rho Z^\rho + Z_{\sigma\lambda\rho}
  Z^{\rho\lambda\sigma} \bigg]
\end{eqnarray}
where
\begin{equation}
  Z^{\mu\nu\rho} = h^\mu_\sigma h^\nu_\tau h^\rho_\lambda \!
  \stackrel{*}{K}{\!}^{\sigma\tau\lambda}
\end{equation}
is the spatial part of the contortion tensor of the semi-teleparallel
geometry and $Z^\mu = Z^{\mu\rho}{}_\rho$ is its trace.  The coupling
constants $c_\sigma$ and $c_\theta$ in \eqref{eq:st_action} are given
by
\begin{align}
  c_\sigma & = \frac{\tilde{c}_\sigma}{\tilde{c}_\sigma + 1} ,\\
  c_\theta & = \frac{2\tilde{c}_\theta}{2 - \tilde{c}_\theta} .
\end{align}
The remaining coupling constants $c_\omega$, $c_{\omega\kappa}$,
$c_\kappa$, $c_{\omega Z}$, $c_{\kappa Z}$, and $c_Z$ are functions of
the coupling constants $\tilde{c}_\omega$, $\tilde{c}_{\omega\kappa}$,
$\tilde{c}_\kappa$, $\tilde{c}_{\kappa a}$, $\tilde{c}_a$.

The action \eqref{eq:st_action} differs from the original action
\eqref{eq:action} in the presence of terms quadratic in the spatial
torsion and coupling terms of the acceleration, vorticity, and spin
rotation with the trace of the spatial torsion. Furthermore, there are
no terms of the form $c_{\kappa a} \kappa\cdot a$ and $c_a a^2$ in the
action. Nevertheless, in the source-free case, the dynamics following
from action \eqref{eq:st_action} is effectively the same as the one
following from the action \eqref{eq:action}. To see this, we first
note that --- analogously to the tensor $Q_{\mu\nu\rho}$ in Equation
\eqref{eq:q} --- the spatial contortion tensor $Z_{\mu\nu\rho}$ can be
decomposed into irreducible parts according to
\begin{equation}
  \label{eq:z}
  Z_{\mu\nu\rho} = Z_{[\mu} h_{\nu]\rho} - \varepsilon_{\mu\nu\sigma}
  Z^\sigma{}_\rho - \frac{1}{3} \varepsilon_{\mu\nu\rho} Z .
\end{equation}  
where $Z$ is a scalar and $Z_{\mu\nu}$ a symmetric trace-free
tensor. Variation of the action \eqref{eq:st_action} with respect to
the semi-teleparallel connection $\stackrel{*}{\omega}{\!}^a{}_{b\mu}$
leads to the field equations
\begin{equation}
  Z = 0, \qquad
  Z_\mu= - \frac{1}{1+2c_Z} \left( 2 a_\mu  + c_{\omega Z} \omega_\mu  +
  c_{\kappa Z} \kappa_\mu \right), \qquad
  Z_{\mu\nu} = 0 .
\end{equation}
Inserting these field equation back into the action
\eqref{eq:st_action} leads to an action of the form \eqref{eq:action}.

Although the coupling constants in the action \eqref{eq:st_action} are
by construction not independent, the action can be used as a starting
point for a semi-teleparallel theory of gravity with independent
coupling constants. Moreover, it is possible to add terms of the form
$\theta Z$ and $\sigma_{\mu\nu} Z^{\mu\nu}$, which couple the spatial
torsion with the expansion and the shear, as well as a further term
quadratic in the spatial torsion of the form
$Z_{\mu\nu\rho}Z^{\mu\nu\rho}$.  Since
$\stackrel{*}{K}{\!}_{\mu\nu\rho} = Z_{\mu\nu\rho} + S_{\mu\nu\rho}$,
the general action has the form
\begin{equation}
 {S}[e_a{}^\mu,\stackrel{*}{\omega}{\!}^a{}_{b\mu}] = \frac{1}{16\pi
   G} \int d^4x \left( R + {\cal K}^{\kappa\lambda\mu\rho\nu\sigma}
 \stackrel{*}{K}{\!}_{\kappa\nu\mu}
 \stackrel{*}{K}{\!}_{\lambda\sigma\rho} \right)
\end{equation}
with a suitable choice of the supermetric ${\cal
  K}^{\kappa\lambda\mu\rho\nu\sigma}$.

%%%%%%%%%%%%%%%%%%%%%%%%%%%%%%%%%%%%%%%%%%%%%%%%%%%%%%%%%%%%%%%%%%%%%%%%%

\section{Conclusions}
In this paper, an extension of Einstein-aether theory has been
proposed which incorporates internal rotational degrees of freedom of
the aether. The main objectives of the paper are

(1) The introduction of the spin rotation $\kappa_{\mu\nu}$ as an
additional kinematic quantity.  The fact that this tensor naturally
appears besides the acceleration, vorticity, shear, and expansion in
the time-space decomposition of the Ricci rotation coefficients
gives this approach a certain completeness.

(2) The formulation of a theory of gravitation that incorporates
couplings of the spin rotation with the vorticity and the
acceleration.

(3) The study of the theory in the weak field limit.  In the case of
dynamical fields, the spin rotation, the vorticity, and the
acceleration are linearly related which allows to eliminate one of
them from the field equations. The field equations acquire the
simplest form if the spin rotation is eliminated. In that case, the
linearized theory has the form of Einstein-aether theory with rescaled
coupling constants.

(4) The formulation of the theory as a (semi-)teleparallel theory of
gravitation. The geometry in this approach is adapted to the
symmetries of the spinning aether. As a result, the kinematic
quantities are part of the torsion tensor. In the matter-free case,
the spatial torsion is algebraically related to the kinematic
quantities which makes the approach effectively equivalent to the
Riemannian formulation. However, if matter fields are present, the
semi-teleparallel formulation may be different from the Riemannian
one.

There are many open questions connected with the Einstein-spin-aether
theory which concern the physical implications of the spin
rotation. These could be elucidated by studying exact solutions of the
nonlinear theory such as black hole solutions and cosmological
solutions. It is to be expected that the simple relation between the
kinematic quantities found in the linearized theory will no longer
hold in solutions of the nonlinear theory. Further open questions
concern the coupling to matter, in particular to spinning matter,
where a semi-teleparallel formulation may lead to different
interactions than the Riemannian formulation, the Hamiltonian
formulation of the extended Einstein-aether theory, and the study of
the observational constraints on the coupling constants.

%%%%%%%%%%%%%%%%%%%%%%%%%%%%%%%%%%%%%%%%%%%%%%%%%%%%%%%%%%%%%%%%%%%%%%%%%

\appendix
\section{Field Equations}

In this appendix we will list the projections $[E_{0i}]$, $[E_{ij}]$,
and $[A_{0i}]$ of the field equations \eqref{eq:fieldeq1},
\eqref{eq:fieldeq2} not given in Section III.

\begin{align}
  \nonumber [E_{0i}] \qquad & \frac{1}{3} \left( \frac{1}{2} c_\theta
  + 2 \right) \left( \partial_\mu \theta - u_\mu \partial_u \theta
  \right) + \left( \frac{1}{2} c_\sigma -1 \right) \left( D_\rho
  \sigma^\rho{}_\mu + u_\mu \sigma^2 + \sigma_{\mu\rho} a^\rho \right)
  - \left( \frac{1}{2} c_\omega - 1 \right) \left( D_\rho
  \omega^\rho{}_\mu + u_\mu \omega^2 + \omega_{\mu\rho} a^\rho \right)
  \\ \nonumber & + c_{\omega\kappa} \left[ \frac{1}{2} D_\rho \left(
    2\omega^\rho{}_\mu - \kappa^\rho{}_\mu \right) + \frac{1}{2} u_\mu
    \omega \cdot \left( 2 \omega - \kappa \right) - \frac{1}{2}
    \kappa_{\mu\rho} a^\rho + \frac{1}{2} Q_{\rho\sigma\mu}
    \omega^{\rho\sigma} \right] + c_\kappa \left( D_\rho
  \kappa^\rho{}_\mu + u_\mu \omega\cdot\kappa + \frac{1}{2}
  Q_{\rho\sigma\mu} \kappa^{\rho\sigma} \right) \\ \nonumber & +
  c_{\kappa a} \left[ D_u \left( 2 \omega_\mu + \kappa_\mu \right) +
    \frac{2}{3} \theta \left( 2 \omega_\mu + \kappa_\mu \right) -
    \sigma_\mu{}^\rho \left( 2 \omega_\rho + \kappa_\rho \right) +
    \frac{1}{2} u_\mu \left( 2 \omega - \kappa \right) \cdot a + 3
    \omega_{\mu\rho} \kappa^{\rho} + \frac{1}{2} Q_{\rho\sigma\mu}
    a^{\rho\sigma} \right] \\ & - c_a \left( D_u a_\mu - \frac{1}{2}
  u_\mu a^2 - \sigma_{\mu\rho} a^\rho + 3 \omega_{\mu\rho} a^\rho +
  \frac{2}{3} a_\mu \theta \right) = 0 \\
  \nonumber [E_{ij}] \qquad & h^\rho_\mu h^\sigma_\nu R_{\rho\sigma} =
  \frac{1}{6} c_\theta h_{\mu\nu} \left( \partial_u \theta + \theta^2
  \right) - c_\sigma \left( D_u \sigma_{\mu\nu} + 2 u_{(\mu}
  \sigma_{\nu)\rho} a^\rho + \sigma_{\mu\nu} \theta +
  2\sigma_{(\mu}{}^\rho \omega_{\nu)\rho} \right)  - 2
  c_\omega \omega_{\rho\mu} \omega^\rho{}_\nu \\ \nonumber & + 2 c_{\omega\kappa}
  \left( \omega_{\rho(\mu} \sigma^\rho{}_{\nu)} + \omega_{\rho\mu}
  \omega^\rho{}_\nu - \omega_{\rho(\mu} \kappa^\rho{}_{\nu)} \right) +
  2 c_\kappa \left( \kappa_{\rho(\mu} \sigma^\rho{}_{\nu)} +
  \kappa_{\rho(\mu} \omega^\rho{}_{\nu)} \right) \\ & + 2 c_{\kappa a}
  \left[ \frac{1}{2} h_{\mu\nu} D_\rho\left( \omega^\rho + \kappa^\rho
    \right) + a_{(\mu} \left( \omega_{\nu)} + \kappa_{\nu)} \right) +
    \sigma_{\rho(\mu} a^\rho{}_{\nu)} \right] - c_a \left( h_{\mu\nu}
  D_\rho a^\rho + 2 a_\mu a_\nu \right) \\
   \nonumber [A_{0i}] \qquad & \frac{1}{6} c_\theta \left( \partial_\mu
  \theta - u_\mu \partial_u \theta \right) + \frac{1}{2} c_\sigma
  \left(D_\rho \sigma^\rho{}_\mu + u_\mu \sigma^2 + \sigma_{\mu\rho}
  a^\rho \right) + \frac{1}{2} c_\omega \left( D_\rho
  \omega^\rho{}_\mu + u_\mu \omega^2 + \omega_{\mu\rho} a^\rho \right)
  \\ \nonumber & + \frac{1}{2} c_{\omega\kappa} \left( D_\rho
  \kappa^\rho{}_\mu + u_\mu \omega\cdot\kappa + \kappa_{\mu\rho}
  a^\rho - 2 \omega_{\mu\rho} a^\rho + Q_{\rho\sigma\mu}
  \omega^{\rho\sigma} \right) - c_\kappa \left( \kappa_{\mu\rho}
  a^\rho - \frac{1}{2} Q_{\rho\sigma\mu} \kappa^{\rho\sigma} \right)
  \\ \nonumber & - c_{\kappa a} \left( D_u \kappa_\mu - \frac{1}{2} u_\mu \kappa\cdot a -
  \sigma_{\mu\rho} \kappa^\rho - \omega_{\mu\rho} \kappa^\rho +
  \frac{2}{3} \kappa_\mu \theta - \frac{1}{2} Q_{\rho\sigma\mu}
  a^{\rho\sigma} \right) \\ & + c_a \left( D_u a_\mu - \frac{1}{2} u_\mu a^2 -
  \sigma_{\mu\rho} a^\rho - \omega_{\mu\rho} a^\rho + \frac{2}{3}
  a_\mu \theta \right) = 0 
\end{align}

\section{Derivation of Equation \eqref{eq:bianchi}}

Starting with the second Bianchi identity,
\begin{equation}
  D_\mu R_{\nu\sigma\rho\tau} + D_\rho R_{\nu\sigma\tau\mu} + D_\tau R_{\nu\sigma\mu\rho} = 0 ,
\end{equation}
contracting with $g^{\sigma\tau}$ leads to
\begin{equation}
\label{eq:contrbianchi}
D_\mu R_{\nu\rho} + D_\sigma R^\sigma{}_{\nu\rho\mu} - D_\rho R_{\mu\nu} = 0 .
\end{equation}
For weak fields, it follows
\begin{equation}
  \label{eq:lincontrbianchi}
 R_{ij,0} = R_{0i,j} + R_{i0j0,0} - R_{j0ik,k}
\end{equation}
where commas denote partial derivatives.  The terms on the right hand
side can be obtained from the Ricci identity \eqref{eq:ricci} in the
following way. Using the first order form of Equation
\eqref{eq:constr2} yields
\begin{equation}
  \label{eq:ricci1}
   R_{0i,j} = - \frac{2}{3} \theta_{,ij} - \sigma_{ki,jk} + \omega_{ki,jk} .
\end{equation}
Furthermore, from Equation \eqref{eq:riemann}, we obtain
\begin{align}
  \label{eq:ricci2}
  R_{i0j0,0} & =  \frac{1}{3} \delta_{ij}\theta_{,00} - \sigma_{ij,00}
  + \omega_{ij,00}  + a_{i,j0}, \\
  \label{eq:ricci3}
  R_{j0ik,k} & = - \frac{1}{3} \theta_{,ij} + \frac{1}{3} \delta_{ij}
  \theta_{kk}\sigma_{kj,ik} - \sigma_{ij,kk} + \omega_{kj,ik} -
  \omega_{ij,kk} .
\end{align}
Inserting Equations \eqref{eq:ricci1} - \eqref{eq:ricci3} into
Equation \eqref{eq:lincontrbianchi} and taking the symmetric part
yields
\begin{equation}
  \label{eq:rdot}
%  R_{ij,0} = a_{(i,j)0} - \sigma_{k(i,j)k} + \sigma_{ij,kk} -
%  \sigma_{ij,00} - \frac{1}{3} \theta_{,ij} - \frac{1}{3} \delta_{ij}
  % \left( \theta_{,kk} - \theta_{,00} \right) .
%  
  \bm{\dot{R}} = \left( \nabla\dot{\bm{a}}\right)_{\textrm{sym}} -
  2 \left(\nabla\left(\nabla\cdot\bm{\sigma}\right)\right)_{\textrm{sym}}
  + \Delta\bm{\sigma} - \ddot{\bm{\sigma}} -
  \frac{1}{3}\nabla\nabla\theta - \frac{1}{3}\bm{1} \left( \Delta
  \theta - \ddot{\theta} \right)
\end{equation}
where vector notation has been used. The time derivative of $\bm{R}$
can also be obtained from the field equation \eqref{eq:dyn_iij}
resulting in
\begin{equation}
  \label{eq:rdot2}
\bm{\dot{R}} = -\frac{1}{3}\left[ \frac{c_\theta}{2} -
  \frac{3\bar{c}_a\left( 1 + \frac{c_\theta}{2} \right)}{1-\bar{c}_a}
  \right] \bm{1} \ddot{\theta} - c_\sigma \bm{\ddot{\sigma}} .
\end{equation}
Equating \eqref{eq:rdot} and \eqref{eq:rdot2} yields Equation
\eqref{eq:bianchi}.

\section{Linearized plane wave solutions}

We assume that the kinematic quantities and the tetrad in first
order approximation are given by plane waves,
\begin{align}
  \label{eq:planewave1}
  S_{\mu\nu\rho} & = \hat{S}_{\mu\nu\rho} e^{ik_\sigma x^\sigma} ,\\
  \label{eq:planewave2}
  \psi_{\mu\nu} & = \hat{\psi}_{\mu\nu} e^{ik_\sigma x^\sigma}
\end{align}
where $k_\mu$ is the wavevector with $(\bm{k})_i = k_i$ and $\bm{k} =
\bm{n} k$.  The spatial vector $\bm{n}$ is the normal of the wave
front and $s=k_0/k$ is its speed. In the following, we will give the
solutions to the field equations which are obtained by insertion of
\eqref{eq:planewave1} and \eqref{eq:planewave2} into the field
equations and Equation \eqref{eq:bianchi}.

\subsection{Spin 0 expansion waves}

We assume that the expansion is given by $\theta = \hat{\theta}
e^{ik_\sigma x^\sigma}$.  The other kinematic quantities are then
given by
\begin{align}
  \bm{a} & = - s_\theta \frac{1+\frac{c_\theta}{2}}{1-\bar{c}_a}
  \theta \bm{n},\\
  \bm{\omega} & = \bm{0}, \\
  \bm{\kappa} & = s_\theta
  \frac{c_{\kappa a}}{c_\kappa}
  \frac{1+\frac{c_\theta}{2}}{1-\bar{c}_a} \theta \bm{n},\\
  \bm{\sigma} & = - \frac{1+\frac{c_\theta}{2}}{1 - c_\sigma} \theta
  \left( \bm{n} \bm{n} - \frac{1}{3} \bm{1} \right) .
\end{align}

Using the tetrad, the spatial trace of the first order metric is given
by $\gamma =\hat{\gamma}e^{ik_\sigma x^\sigma}$. The other components
of $\psi_{\mu\nu}$ in the gauge \eqref{eq:gauge} then are
\begin{align}
  \bm{A} & = \bm{0} , \\
  \phi & = -s^2_\theta \frac{1+\frac{c_\theta}{2}}{1-\bar{c}_a} \gamma , \\
  \bm{\zeta} & = s_\theta \frac{c_{\kappa a}}{c_\kappa}
  \frac{1+\frac{c_\theta}{2}}{1-\bar{c}_a} \gamma \bm{n} , \\
  \bm{\gamma} & =  \frac{1+\frac{c_\theta}{2}}{1 - c_\sigma} \gamma 
  \left( \bm{n} \bm{n} - \frac{1}{3} \bm{1} \right) .
\end{align}

\subsection{Spin 1 acceleration-vorticity waves}

If the acceleration is given by $\bm{a} = \hat{\bm{a}} e^{ik_\sigma
  x^\sigma}$ with the transversality condition $\bm{n}\cdot\bm{a} =
0$, the kinematic quantities are
\begin{align}
  \theta & = 0 ,\\
  \bm{\omega} & = - \frac{1}{2s_{a\omega}} \bm{n} \times \bm{a} ,\\
  \bm{\kappa} & = - \frac{c_{\kappa a}}{c_\kappa} \bm{a} + 
  \frac{c_{\omega\kappa}}{2s_{a\omega}c_\kappa} \bm{n} \times \bm{a} ,\\
  \bm{\sigma} & = \frac{1}{s_{a\omega}\left( 1 - c_\sigma \right)} \left(
  \bm{n} \bm{a} \right)_{\textrm{sym}} .
\end{align}
Analogously to electromagnetic waves, the condition of transversality
reduces the number of independent modes to two.  In terms of the
tetrad, the potential $\bm{A}$ is given by $\bm{A} = \hat{\bm{A}}
e^{ik_\sigma x^\sigma}$ with the gauge condition $\bm{n}\cdot\bm{A} =
0$. The other components of the first order tetrad are then given by
\begin {align}
  \phi & = 0 ,\\
  \bm{\zeta} & =  \frac{1}{2s_{a\omega}}\left( \frac{c_{\omega\kappa}}{c_\kappa} -1
  \right) \bm{n}\times\bm{A} - \frac{c_{\kappa a}}{c_\kappa} \bm{A} ,\\
  \bm{\gamma} & = - \frac{1}{s_{a\omega}\left( 1 - c_\sigma \right)} \left(
  \bm{n} \bm{A} \right)_{\textrm{sym}} ,\\
  \gamma & = 0 .
\end{align}

\subsection{Spin 2 shear waves}

The shear waves are given by $\bm{\sigma} = \hat{\bm{\sigma}}
e^{ik_\sigma x^\sigma}$ with the transversality condition
$\bm{n}\cdot\bm{\sigma}=\bm{0}$ and all other kinematic quantities
vanishing, $\bm{a}=\bm{\omega}=\bm{\kappa}=\bm{0}$, $\theta=0$. Since
$\bm{\sigma}$ has five independent components and the transversality
condition consists of three equations, there are two independent shear
modes.  If the tetrad is used, the solutions are the gravitational
waves of Einstein gravity with $\bm{\gamma} = \hat{\bm{\gamma}}
e^{ik_\sigma x^\sigma}$ and $\bm{n}\cdot\bm{\gamma} = 0$. All other
components of the first order tetrad are zero, $\phi = \gamma = 0$,
$\bm{A} = \bm{\zeta} = \bm{0}$.

%%%%%%%%%%%%%%%%%%%%%%%%%%%%%%%%%%%%%%%%%%%%%%%%%%%%%%%%%%%%%%%%%%%%%%%%%


\begin{thebibliography}{99}

\bibitem{jacobson01} T.~Jacobson and D.~Mattingly, ``{Gravity with a
  dynamical preferred frame},'' {\em Phys. Rev.} D 64, 024028 (2001)

\bibitem{eling06} C.~Eling, T.~Jacobson, and D.~Mattingly,
  ``{Einstein-aether theory}'' {Deserfest}, ed. J.~Liu, M.J.~Duff,
  K.~Stelle and R.P.~Woodward (Singapore: World Scientific) (2006)
  
\bibitem{jacobson07} T.~Jacobson, ``{Einstein-aether gravity: a status
  report}'',{\em Proceedings, Workshop on From quantum to emergent gravity:
  Theory and phenomenology} (QG-Ph): Trieste, Italy, June 11-15, 2007,
  PoS QG-PH, 020 (2007)

\bibitem{halbwachs60} F.~Halbwachs, ``{Lagrangian Formalism for a
  Classical Relativistic Particle Endowed with Internal Structure},''
  {\em Prog. Theor.  Phys.} {\bf 24} (1960) 291

\bibitem{ray82} J.~R.~Ray, L.~L.~Smalley, ``{Spinning fluids in
  general relativity},'' {\em Phys. Rev.} D 26 (1982) 2619
  
\bibitem{moller61} C.~M{\o}ller, ``{Conservation Laws and Absolute
  Parallelism in General Relativity},'' {\em
  K. Dan. Vidensk. Selsk. Mat. Fys. Skr.} {\bf 1} (1961) no.~10,
  1--50.

\bibitem{ellis71} G.~F.~R.~Ellis, in R.~Sachs, ``{General Relativity and
  Cosmology}'' (Proceedings of the 1959 E.~Fermi Summer School, Varenna,
  XLVII. Course) (London, New York: Academic Press, 1971)

\bibitem{jacobson04} T.~Jacobson and D.~Mattingly, ``{Einstein-Aether
  waves,}'' {\em Phys. Rev.} D 70, 024003 (2004)

\bibitem{gasperini87} M.~Gasperini,''{Singularity prevention and
  broken Lorentz symmetry}'' {\em Class. Quantum Grav.} {\bf 4}
  (1987), 485-494

\bibitem{kohler00} C.~Kohler, ``{Semi-Teleparallel Theories of
  Gravitation}'', {\em Gen. Rel. Grav.} 32 (2000), 1301-1317

\bibitem{capozziello01} S.~Capozziello, G.~Lambiase, and C.~Stornaiolo,
  ``Geometric classification of the torsion tensor of space-time'',
  {\em Ann. Phys.} {\bf 10} (2001) 713-727

\end{thebibliography}
\end{document}